\documentclass{elsart}
\usepackage{graphicx,epsfig}
\usepackage{amssymb}
\begin{document}
\begin{frontmatter}

\title
{Is the Coulomb sum rule violated in nuclei?}
\author[Saclay]{J.~Morgenstern}
\author[Temple]{Z.-E.~Meziani}

\address[Saclay]{CEA Saclay DSM/DAPNIA/SPhN, F91191, Gif-sur-Yvette Cedex,
 France.}
\address[Temple]{Temple University, Philadelphia, PA 19122.}
%\date{\today}
\maketitle
\begin{abstract}
Guided by the  experimental confirmation of the validity of the
Effective Momentum Approximation (EMA) in quasi-elastic scattering off
nuclei, we have re-examined the extraction of the longitudinal and
transverse response functions in medium-weight and heavy nuclei.
In the EMA we have performed a Rosenbluth separation of the  available world
 data on $^{40}$Ca, $^{48}$Ca, $^{56}$Fe, $^{197}$Au, $^{208}$Pb and $^{238}$U.
We find that the longitudinal response function for these nuclei is ``quenched"
and that the Coulomb sum is not saturated, at odds with  claims in the literature.
\end{abstract}
\end{frontmatter}

\newpage

One of the important questions in nuclear physics is how nucleon 
properties are affected by the nuclear medium, since it might form a bridge 
between the strong interaction between nucleons and the underlying  theory 
of Quantum ChromoDynamics (QCD). A good example is the partial restoration 
of chiral symmetry in nuclear matter and its consequence for nucleon properties 
in the nuclear medium (for comprehensive reviews see~\cite{Birse:94,Brown:01}).
Since elastic scattering from a free nucleon has been well measured,
 quasi-elastic electron scattering off nuclei is considered a promising tool 
to investigate the properties of nucleons in nuclei.  In particular, it was 
proposed~\cite{McVoy:62} that  a Rosenbluth separation of the electric and magnetic
responses of a nucleus ($R_L$ and $R_T$, respectively) could test a model-independent
property known as the Coulomb sum rule (CSR). This sum rule states that when integrating
the quasi-elastic $R_L(q,\omega)$ over the full range of energy loss $\omega$ at 
large enough three-momentum transfer $\vert {\mathbf{q}}\vert = q $  (greater than twice
the  Fermi momentum, $q \gtrsim 500$ MeV/c), one should count the number of protons (Z)
in a nucleus. More explicitly the quantity $S_L(q)$  defined by
\begin{eqnarray}
S_L(q) = {\frac{1}{Z}}\int_{0^+}^{\infty}
 \frac{R_L(q,\omega)}{\tilde{G_E}^2}d\omega
\end{eqnarray}
is predicted to be unity in the limit of large $q$. Here  ${\tilde{G_E}} = 
(G_E^p + N/Z G_E^n) \zeta$ takes into account the nucleon charge form factor 
inside the nucleus (which is usually taken to be equal to that of a free nucleon) 
as well as a relativistic correction ($\zeta$) suggested by de Forest~\cite{deForest:84}. 
The lower limit of integration $0^+$ excludes the elastic peak.

This simple picture can be polluted by the modification of the free nucleon 
electromagnetic properties by the nuclear medium and the presence of nucleon-nucleon 
short-range correlations. There is general agreement that around $q$ of 500~MeV/c,
$S_L$ should not deviate more than a few percent from unity due to
nucleon-nucleon correlations, and reach unity at higher $q$-values,
independent of the nucleon-nucleon force chosen (see the review paper~\cite{Orlandini:91}). 
Thus, a result of $S_L$ far from unity might indicate a modification of 
the nucleon electric properties in the nuclear medium.

In the last twenty years a large experimental program has been carried out at
Bates~\cite{Altemus:80,Deady:83,Hotta:84,Deady:86,Blatchley:86,Dytman:88,Dow:88,Yates:93,Williamson:97},
Saclay~\cite{Barreau:83,Meziani:84,Meziani:85,Marchand:85,Zghiche:94} and
SLAC~\cite{Baran:88,Chen:91,Meziani:92} aimed at the  extraction of
$R_L$ and $R_T$ for a variety of nuclei. Unfortunately, in
the case of medium-weight and heavy nuclei conclusions reached by different
 experiments ranged from a full saturation of the CSR to its violation by 30~\%.
As a result a spectrum of explanations has emerged ranging from questioning the
validity of the experiments (i.e., experimental backgrounds), 
inadequate Coulomb corrections (especially for heavy
nuclei) to suggesting a picture of a ``swollen nucleon" in the nuclear medium due to a partial
 deconfinement~\cite{Noble:81,Celenza:84,Mulders:86,Ericson:86,Brown:89}.

Up to now the Coulomb corrections for inclusive experiments have been
evaluated theoretically by two independent groups, one from Trento
 University~\cite{Traini:88,Traini:95,Traini:01} and the other from Ohio University
~\cite{Wright:92}.  The Trento group found that the Effective 
Momentum Approximation (EMA) works with an accuracy better than 1\%, while 
the Ohio group derived significant corrections beyond EMA. All useful quantities for the EMA
are defined in~\cite{Traini:88,Traini:95,Rosen:80,Gueye:99}. A detailed
discussion of the different theoretical approaches can be found
 in~\cite{Traini:01}.  Previous extractions of $R_L$ and $R_T$ were 
performed either without Coulomb corrections in~\cite{Meziani:84,Meziani:85} or  
by applying the Trento group calculations ~\cite{Zghiche:94}, or the
Ohio group calculations~\cite{Williamson:97,Jourdan:95}.
This led to questionable results even when Coulomb corrections from either 
groups were applied, particularly in the region beyond the quasielastic peak 
known as the "dip region" since meson exchange currents and pion production 
while significant, but were not included in any of the nuclear models used.

Recently and for the first time the Coulomb corrections have been studied
in a direct comparison of quasielastic electron and positron scattering off 
$^{12}$C and $^{208}$Pb at forward (Fig.~1a) and backward (Fig.~1b) 
angles~\cite{Gueye:99}. It has been found experimentally that the EMA can 
adequately describe the electron and positron
scattering over the entire quasielastic and dip regions. For the
 quasielastic region were theoretical calculations have been performed, this 
comparison is in agreement with Traini and collaborators'
result~\cite{Traini:88,Traini:95,Traini:01} and in disagreement
with the Ohio group's result as shown in Fig.~1. Recent full DWBA calculations
of the Ohio group~\cite{Kim:01} are presented here instead of the LEMA calculations
 presented in ~\cite{Gueye:99}, nevertheless, the disagreement with the experimental 
comparison persists.
 Values of the effective Coulomb potential $\widetilde{V_C}$, equal to half the
 difference between the electrons and positrons incident energies,
were extracted from this comparison allowing us to separate $R_L$ and $R_T$ with
 the EMA independently from any theoretical calculations of the Coulomb corrections.
The values of $\widetilde{V_C}$ were found to be very close to the average
Coulomb potential of the nucleus and not to the value $V_C(0)$ at the center
of the nucleus (see Table II of Ref.~\cite{Gueye:99}) as used
previously by several authors including ourselves
\cite{Blatchley:86,Williamson:97,Zghiche:94,Jourdan:95}.

We present here the results of a re-analysis of the Saclay data only
using the Coulomb corrections {\it based on the EMA}
to extract $R_L$ and $R_T$ and evaluate $S_L(q)$. Our goal was to
first determine the change in our previously reported results which either
had no Coulomb corrections applied, for $^{40}$Ca, $^{48}$Ca and 
$^{56}$Fe~\cite{Meziani:84} or for $^{208}$Pb~\cite{Zghiche:94}, had Coulomb corrections 
applied following a procedure described by Traini {\it et al.}~\cite{Traini:88} 
with $V_C(0)$ instead of  $\widetilde{V_C}$ and a too crude nuclear model
which generated spurious higher order corrections. Next, it was important to
test whether the data from SLAC and Bates analyzed within the EMA would influence 
our original results as quoted in ~\cite{Jourdan:95} for the case of $^{56}$Fe.
For that purpose, we present the results obtained with the EMA by combining
data from Saclay, Bates and SLAC on $^{40}$Ca, $^{48}$Ca, $^{56}$Fe, $^{197}$Au,
$^{208}$Pb and 
$^{238}$U~\cite{Hotta:84,Deady:86,Williamson:97,Meziani:84,Zghiche:94,Baran:88,Day:93}.
In order to combine different nuclei at the same kinematics, we normalized
each nucleus with the factor $\textbf{K}=Z[(\epsilon\sigma_{ep}^L+\sigma_{ep}^T)
+N(\epsilon\sigma_{en}^L+\sigma_{en}^T)]$,  where $\epsilon$ is the virtual 
photon polarization and  $\sigma_{ep(n)}^{L(T)}$ is the longitudinal (transverse) 
virtual photon-proton (-neutron) cross section.   We conclude by evaluating $S_L$ 
and testing the  Coulomb sum rule.

In Fig.~2 we present the results of the Rosenbluth separation  at  
$q_{eff}$ = 570 MeV/c, the same $q_{eff}$ as used in Jourdan's analysis.
 In our original publication the highest $q$-value chosen was 550 MeV/c to avoid 
regions of high $\omega$ where systematic errors are large and difficult to 
estimate. There is a clear disagreement between the results in \cite{Jourdan:95}
 and the present analysis above $\omega$ = 150 MeV for $R_T$ and 
$\omega$ = 230 MeV for $R_L$. The difference between these results is 
significant for both $R_L$ and $R_T$ and we attribute it to the Coulomb  corrections used
in ~\cite{Jourdan:95} following the Ohio group
calculations~\cite{Wright:92} since, as shown in Fig.~1, these corrections do
not reproduce the EMA behavior observed in the comparison of electron
and positron quasielastic cross section~\cite{Gueye:99}. 
Within the EMA, the same nominal momentum is obtained by adding at each incident energy
a constant negative value $\widetilde V_C$. Therefore, larger cross sections are used 
to perform the Rosenbluth separation of $R_L$ and $R_T$ because the new incident energies 
are lower at all angles. However, due to the lower value of the incident energy required at
backward angles for the same $q_{eff}$ , the relative increase in the  cross
section is more sizeable at backward angles than at forward angles. Consequently, within EMA,
$R_T$ is increased and $R_L$ decreased.  This effect was previously seen
in the results of SLAC experiment  NE9~\cite{Chen:91} at $q_{eff}$ = 1 GeV/c. However,  as shown
in Fig.~2,  the Coulomb corrections applied in~\cite{Jourdan:95} following the prescription 
described in ~\cite{Wright:92} have the opposite effect, namely to decrease $R_T$ 
and to enhance $R_L$. We note that the results of the present analysis 
are only slightly changed when we combine the
forward-angle SLAC NE3~\cite{Day:93} and the Saclay data.

The situation for the Bates measurements on $^{40}$Ca~\cite{Williamson:97} and
$^{238}$U~\cite{Blatchley:86}
requires further clarification. Backward-angle cross sections were measured in an early stage
of the experiment, where secondary scattering background was present. This
background was estimated in part by performing some experimental tests
and corrected using a simulation code. Forward-angle cross sections,
$^{238}$U at $60^{\circ}$ and $^{40}$Ca at $45.5^{\circ}$, have been measured
 with a modified experimental setup. Cross sections of $^{56}$Fe at
$180^{\circ}$~\cite{Hotta:84}  have been also measured at Bates with another setup.
 In Fig.~3 we have  compared backward-angle data by comparing the transverse responses. 
The $^{56}$Fe $180^{\circ}$ data is
 purely transverse, and transverse responses obtained after separation 
from $^{56}$Fe measurements at $140^{\circ}$, $143^{\circ}$, $160^{\circ}$, 
depend very little on the uncertainties
of the forward-angle measurements.  We can observe a good  agreement
between Saclay and the $180^{\circ}$ $^{56}$Fe measurements from Bates. However,
discrepancies between the $^{40}$Ca  backward angles data from Bates and Saclay
 (Fig.~3a), and $^{238}$U from Bates and $^{208}$Pb from Saclay (Fig.~3b) are 
observed. Part of these discrepancies are due to the Coulomb corrections, but 
there remain experimental differences in spite of the background corrections 
performed in the Bates experiments.
Fig.~3c shows the total responses at $60^{\circ}$ of $^{238}$U from Bates with
 the new setup and of $^{208}$Pb from Saclay in fairly good agreement. Also, 
longitudinal and transverse responses of $^4$He and $^3$He obtained in a Rosenbluth 
separation from forward and backward angle measurements
 using the new experimental setup at Bates~\cite{Dytman:88,Dow:88}
are also in good agreement with the Saclay response
 functions~\cite{Marchand:85,Zghiche:94} as shown in Fig.~4.

In Fig.~5 we show results for $R_L$ (a) and $R_T$ (b) at $q_{eff}$ = 550 MeV/c
 and for $R_L$ (c) at 500 MeV/c  obtained with a re-analysis of $^{208}$Pb in the
 EMA~\cite{Gueye:99}. The data are compared to the previously published work of 
Zghiche {\it et al.}~\cite{Zghiche:94}.
 Furthermore, for a consistency check of our analysis, we present in Fig.~5 results 
obtained by combining the Saclay data, the $^{197}$Au SLAC
 data~\cite{Day:93} and the $^{238}$U Bates data~\cite{Blatchley:86}. Both data
 sets were renormalized to $^{208}$Pb with the factor \textbf{K}, equal to 1.05 
for $^{197}$Au and 0.88 for $^{238}$U. For $^{238}$U we have used only the 
60$^{\circ}$ data taken with the new experimental setup but not data at backward 
angle taken with the earlier setup~\cite{Blatchley:99}.
While there is a clear difference in $R_T$ between the previously published
 work~\cite{Zghiche:94} and this analysis the conclusions regarding the quenching of
$R_L$ have not  changed qualitatively. Figure~5 also shows that combining the SLAC,
Bates and Saclay data to extract $R_L$ and $R_T$ does not change the results
significantly. We also present in Fig.~5 (b and c panels)
microscopic Nuclear Matter calculations (NM) of $R_L$ at 550 and 500
MeV/c~\cite{Fabrocini:89}  (dashed lines) and 
Hartree-Fock calculations (HF) of $R_L$  at 500 MeV/c including short-range
correlations and final-state interaction~\cite{Traini:93}
(solid line). If the integrated strengths of $R_L$ within the experimental
 limits are quite close (5\% more strength for HF; see Fig.~6, compared to NM), the
shapes are different. The large energy excitation tail of $R_L$ 
is much less important in the HF than in the NM calculation. Finally, we
plotted in Fig.5c the HF calculation with a modified form
factor~\cite{Soyeur:93,Traini:86} (dotted-dashed curve) (discussed later in the text) and
find a fairly good agreement with the combined Saclay and Bates-60$^{\circ}$ data
(triangles down).

We now turn to the results of the experimental Coulomb sum but first
discuss the quantitative difference between the EMA analysis
and that of Ref.~\cite{Jourdan:95} as summarized in Table 1.
A comparison between the present result of $S_L$ for $^{56}$Fe and that of
Ref.~\cite{Jourdan:95} identifies two possible sources for the difference; (a)
the Coulomb corrections and (b) the use of the total error in the Saclay data
but only the statistical error in the SLAC data. For (a), we believe that
the Coulomb corrections used in~\cite{Jourdan:95}
following the prescription of~\cite{Wright:92}, at variance with
the experimental confirmation of the EMA~\cite{Gueye:99}, have the wrong sign;
they increase the longitudinal response instead of decreasing it. The Coulomb
corrections within the EMA reduce $S_L$ by 10\% while it is increased 
by 5\% in~\cite{Jourdan:95} .
 For (b), more weight was given to the SLAC NE3 data by
neglecting the 3.5\% systematic error quoted by the authors~\cite{Day:93},
leading to an artificial enhancement of $R_L$ by 4$\%$.

Fig.~6 shows the results  obtained in the present analysis for $S_L$ of $^{40}$Ca,
$^{48}$Ca, $^{56}$Fe and $^{208}$Pb. In Fig.~6a the data shown 
were obtained using only cross section measured at Saclay, whereas in Fig.~6b the results 
by combining data from at least two different laboratories among Bates,
Saclay and SLAC except for the data point from SLAC experiment NE9 at $q_{eff}$=1.14~GeV/c.
Among the Bates cross-section data of $^{40}$Ca and $^{238}$U we chose to use only those
measured at forward angles  with the modified experimental setup. In order to evaluate
$S_L$ we used the Simon~\cite{Simon:80} parametrization of the proton charge form
factor, while for the neutron charge form factor we have taken into account the
 data by Herberg {\it et al.}~\cite{Herberg:99}. We note that for $^{208}$Pb
the total error in the experimental determination of $\widetilde{V_C}$ is
1.5~MeV leading to a relative uncertainty of 2~\% on
$S_L$ at $q_{eff}=$~500 MeV/c. We have plotted the total NM
Coulomb sum~\cite{Schiavilla:87} (solid line), a partial NM Coulomb sum
integrated only within the experimental limits at 400 $\leq q_{eff} \leq$ 550
MeV/c~\cite{Fabrocini:89} (dashed curve) and a partial HF Coulomb sum in $^{208}$Pb
integrated within the experimental limits at $q_{eff}$=500 MeV/c ~\cite{Traini:93}
(thick right cross).  The experimental results are to be compared with the partial sum
and not the total sum values.  We observe a quenching between 20$\%$ and 30$\%$  in all
medium and heavy nuclei.

The observed quenching is similar to the quenching of the ratio $R_L$/$R_T$
 observed in a $^{40}$Ca~(e,e'p)~$^{39}$K  experiment~\cite{Reffay:88} which was
performed at energy transfers $\omega$ near or below the maximum of the quasi-elastic
peak ($\omega$ $\lesssim \omega_{max}$) where the quasi-elastic process is dominant.
The observed quenching of $R_L$/$R_T$ implies that $R_T$ is little affected
by the medium while $R_L$ is reduced.  On the other hand, when analyzing
the SLAC data~\cite{Day:93,Sick:85}, it has been observed that the unseparated
 cross sections scale at momentum transfers q$\gtrsim$ 2~GeV/c for $\omega$ $\lesssim
 \omega_{max}$. It was pointed out in~\cite{Sick:85} that 
 this scaling is destroyed if one introduces medium effects
in the nucleon form factors. However, at these momentum transfers
the longitudinal component represents  only 20$\%$ or less of the total cross
 section; a quenching of  the longitudinal response ranging from 20$\%$ to 30$\%$ produces
a quenching between 4$\%$ and 6$\%$ for the  unseparated cross sections,
which clearly remains within the experimental band of the scaling
 representation~\cite{Meziani:86}.
 Consequently, the conclusion that no  medium effects are observed
applies essentially to the transverse response, in agreement with what
we obtain from the Saclay (e, e')  and (e, e'p) experiments.

 Several authors have proposed models for medium effects to explain this
 quenching~\cite{Noble:81,Celenza:84,Mulders:86},
 but found it difficult to explain why only  $R_L$ was affected by the medium.
A later model based on chiral-symmetry restoration in
 nuclei~\cite{Brown:89,Soyeur:93}
predicted a decrease of vector-meson masses (and consequently a decrease of the
nucleon form factor) inside nuclei. In this model only  $R_L$ is affected while
 $R_T$ changes very little because the magnetic operator is changed by about
the same amount as the magnetic form factor due to the change of the nucleon
free mass into the effective mass. The dot-dashed curve and the thin right cross
 are from  similar calculations to those of the dashed curve and the thick right
 cross except that we have replaced the free nucleon form factor by a modified 
form factor in
$^{208}$Pb calculated in Ref.~\cite{Soyeur:93}. We can see that there is a good agreement with
the data. A quenching of about 20\% of $R_L$  with a small change of $R_T$ has also been
 predicted in  calculations based on an improved Walecka model~\cite{Serot:86}
 using density dependent coupling constants and relativistic RPA
correlations~\cite{Caillon:95,Saito:99}.

In conclusion, there is a  good agreement between the data from Saclay,
SLAC, Bates 180$^{\circ}$ experiments and Bates data taken with the new
setup. We believe that we have established experimentally the existence of
a quenching of $S_L$ in medium and heavy nuclei as shown in Fig.~6. This
quenching is not observed in low-density nuclei such as $^3$He and
$^2$D~\cite{Marchand:85,Dytman:88} and short-range
correlations are not able to explain this effect. We interpret this as an
indication for a  change of the nucleon properties inside the nuclear medium. If
 we assume the dipole expression for the charge form factor, the observed
quenching of the CSR would correspond to a relative change of the proton charge
radius of $13\pm4$\% in a  heavy nucleus. The accuracy of the CSR could  be
 improved and the $q$ region extended up to 1~GeV/c with  the new generation 
of electron accelerators. Such a proposal has been approved recently at
Jefferson Lab~\cite{Choi:01}.

This work was supported by Department of Energy contract DE-FG02-94ER40844
(Z.-E M.) and the Commissariat \`a l'Energie Atomique (J. M.).

\newpage

\begin{table}[top]
%\begin{tabular}{lllll}
\begin{tabular}{|c|c|c|c|c|c|}
\cline{1-6}
Analysis & Saclay & SLAC & SLAC & Coulomb & S$_L$ \\
 &        uncertainty & data &uncertainty & corrections& \\
\cline{1-6}
Jourdan & total & included &statistical & No & 0.86$\pm$0.12\\
        & total & included &statistical & Yes & 0.91$\pm$0.12\\
\cline{1-6}
        Present & total & not included & -     & No & 0.72$\pm$0.23\\
       work    & total  & not included & - & Yes & 0.63$\pm$0.20\\
            & total  & included &total & No & 0.82$\pm$0.12\\
            & total  & included &total & Yes & 0.73$\pm$0.12\\
\cline{1-6}
\end{tabular}
\vskip5mm
\caption{ Comparison of the Coulomb sum results in $^{56}$Fe between Jourdan's
 work and the present analysis.
Total refers to the statistical and systematic uncertainties added in
 quadrature. Jourdan's Coulomb corrections
are described in \protect\cite{Jourdan:95}
following the Ohio group prescription~\protect\cite{Wright:92}.
This work Coulomb corrections when applied are performed following the
 EMA~\protect\cite{Gueye:99}.}
\end{table}

\newpage

\begin{figure}
Figure~1.~e$^+$ (filled circles) and e$^-$ (open circles) total response
 functions at the same effective incident energies along with the Ohio group 
calculations (e$^+$ thick solid lines,
e$^-$ thick dashed lines) and the Trento group calculations (e$^+$ thin solid
 lines,
e$^-$ thin dashed lines). 
\end{figure}

\begin{figure}
Figure~2.~$R_L$ and $R_T$ response functions of $^{56}$Fe extracted at
$q_{eff}$ = 570 MeV/c in the present analysis using
the Saclay data only (circles), then with adding the SLAC data from
NE3~{\protect\cite{Day:93}} (triangles) and from Jourdan's
 analysis~{\protect\cite{Jourdan:95}}
(squares). The result of the original Saclay analysis without Coulomb
 corrections~\protect\cite{Meziani:84,Meziani:85} is indicated by the solid
 line.
\end{figure}

\begin{figure}
Figure~3.~a) Transverse response functions of $^{40}$Ca: Saclay data
(open circles),
Bates results~{\protect\cite{Williamson:97}} (filled triangles), our analysis of Bates data using
EMA (open triangles) and $^{56}$Fe Saclay data (crosses) , Bates data at 180$^{\circ}$ (filled
squares); b) Transverse response functions of $^{208}$Pb: Saclay data (open
circles),
$^{238}$U Bates results (filled triangles), our analysis of Bates results using EMA (open
triangles) and
$^{56}$Fe Bates data at 180$^{\circ}$ for comparison (filled squares).
c) Total response function at 60$^{\circ}$ of $^{208}$Pb (open circles) and
$^{238}$U (filled triangles).
\end{figure}

 \begin{figure}
Figure~4.~Longitudinal and transverse response functions of $^{3}$He and
 $^{4}$He at $q$ = 500 MeV/c. Bates data~{\protect\cite{Dytman:88,Dow:88}}  are the 
open circles and Saclay data~{\protect\cite{Marchand:85,Zghiche:94}}
 are the filled circles.
\end{figure}

\begin{figure}
Figure~5.~Longitudinal (a) and transverse (b) response functions
of $^{208}$Pb at $q_{eff}$ = 550 MeV/c extracted in the EMA. Saclay data 
only (filled circles), combined with
$^{197}$Au-15$^{\circ}$ SLAC data (triangles up), combined with 
Bates$^{238}$U-60$^{\circ}$ data (triangles
down); previous Saclay results with Coulomb corrections~{\protect\cite{Zghiche:94}}:
thin solid lines. c) Longitudinal response function at $q_{eff}$ = 500 MeV/c 
(same experimental symbols). Nuclear matter calculations {\protect\cite{Fabrocini:89}}: dashed
line, Hartree Fock calculations  including short range correlations and final state
interactions~{\protect\cite{Traini:93}} with free nucleon form factors (solid line),
 with modified nucleon form factors (dotted-dashed line).
\end{figure}

\begin{figure}
Figure~6.~$S_L$ obtained in the EMA as a function of $q_{eff}$ using only
Saclay data (a) and using Saclay data
combined with SLAC NE3 and Bates data with the new experimental setup (b).
N-M calculations~{\protect\cite{Fabrocini:89}} (solid line),
N-M calculations integrated within the experimental limits: dashed line,
same with modified form factors (dotted-dashed line),
$^{208}$Pb~H-F calculations~{\protect\cite{Traini:93}} integrated
within the experimental limits (thick right cross),
same with modified form factors (thin right cross).
$^{56}$Fe  SLAC NE9~\protect\cite{Chen:91} (filled circle) and
Jourdan analysis of $^{56}$Fe Saclay data (thick star) are shown in (b).

\end{figure}

\newpage

\begin{figure}
\begin{center}
\centerline{\includegraphics[scale=1.]{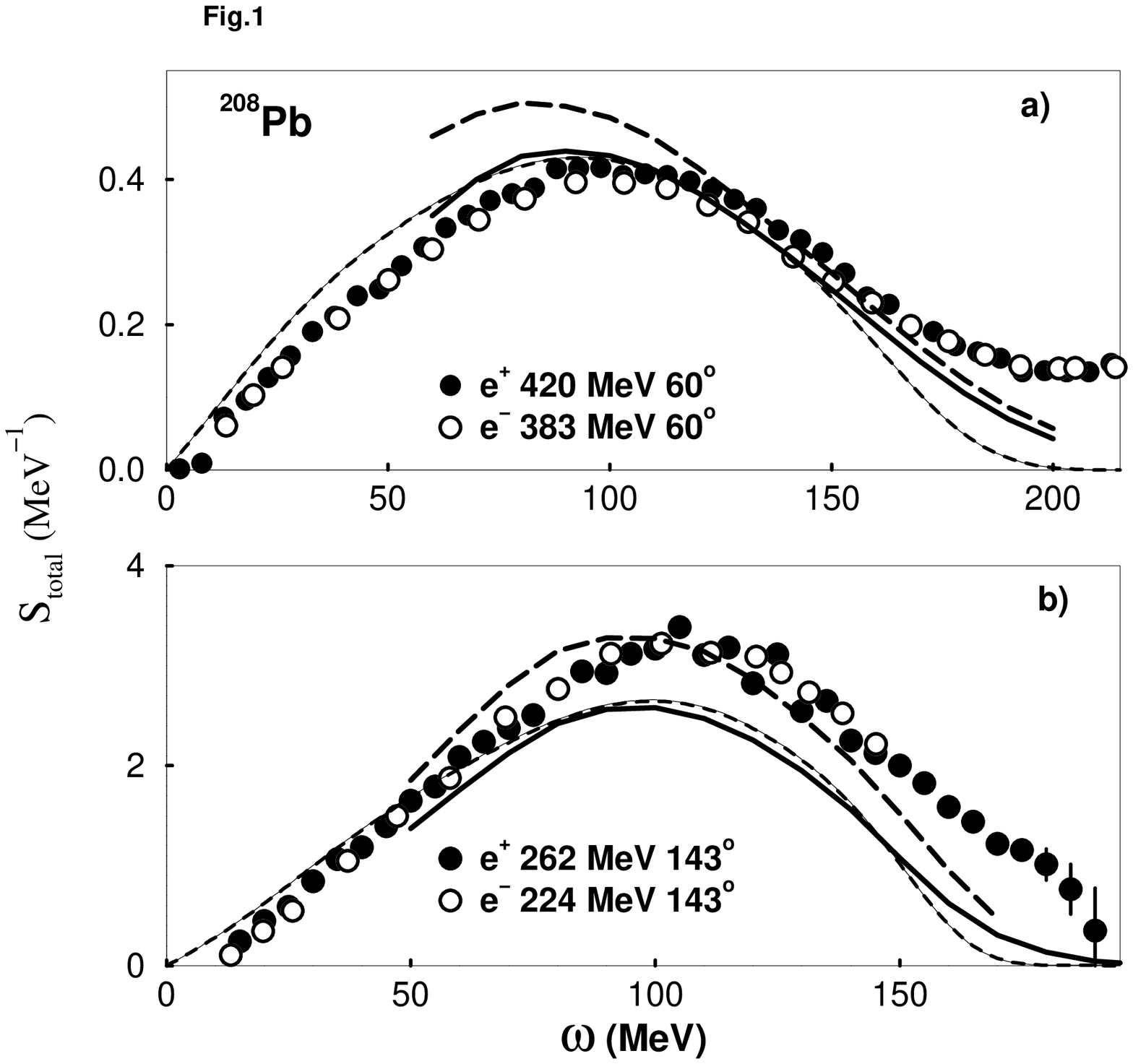}}
\end{center}
\end{figure}

\newpage
\begin{figure}
\begin{center}
%\centerline{\includegraphics[scale=1.]{fig-2-noir.eps}}
\epsfig{file=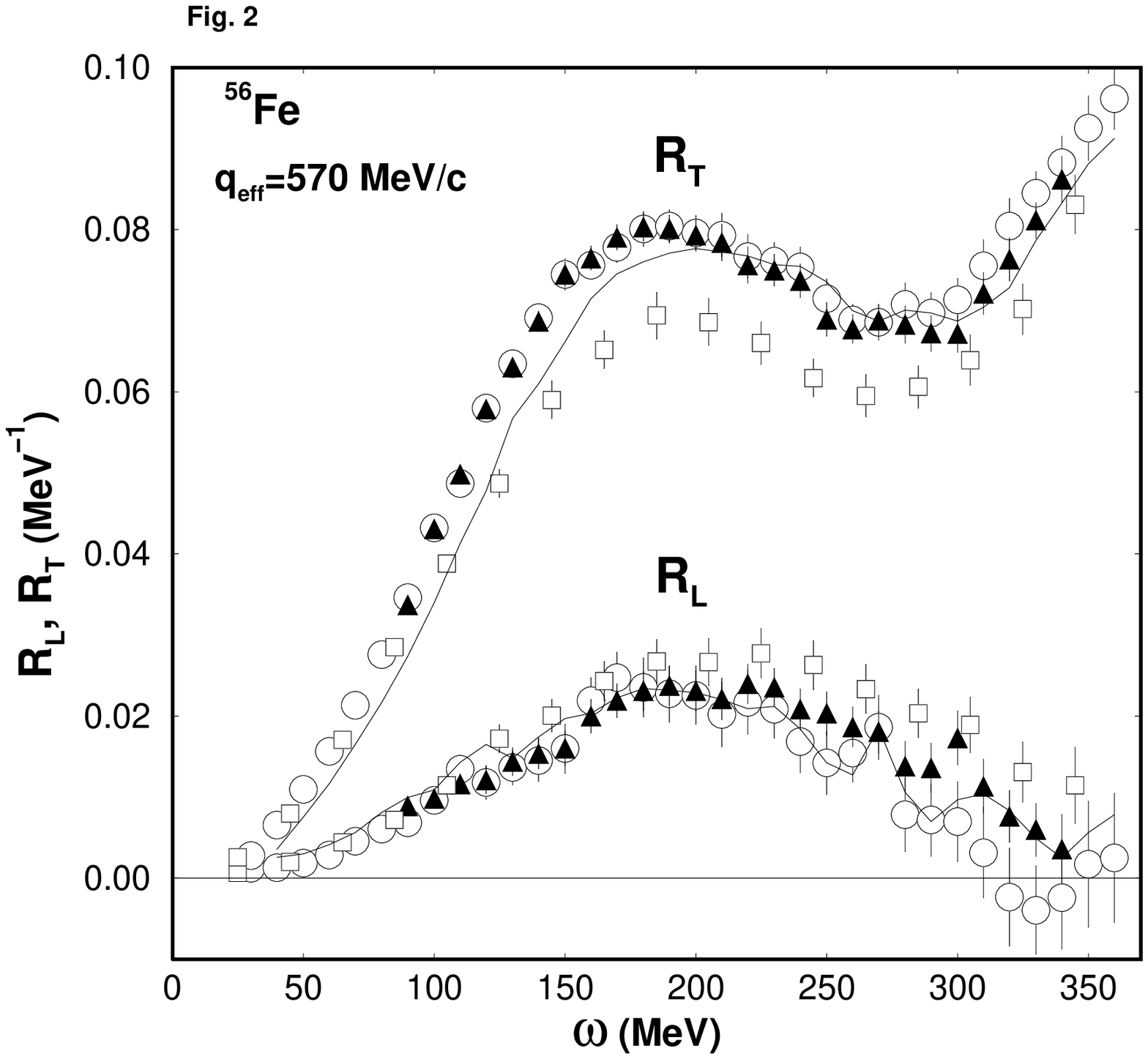,width=18.cm,height=18.cm}
\end{center}
\end{figure}
\newpage

\begin{figure}
\begin{center}
%\centerline{\includegraphics[scale=1]{fig-3-noir.eps}}
\epsfig{file=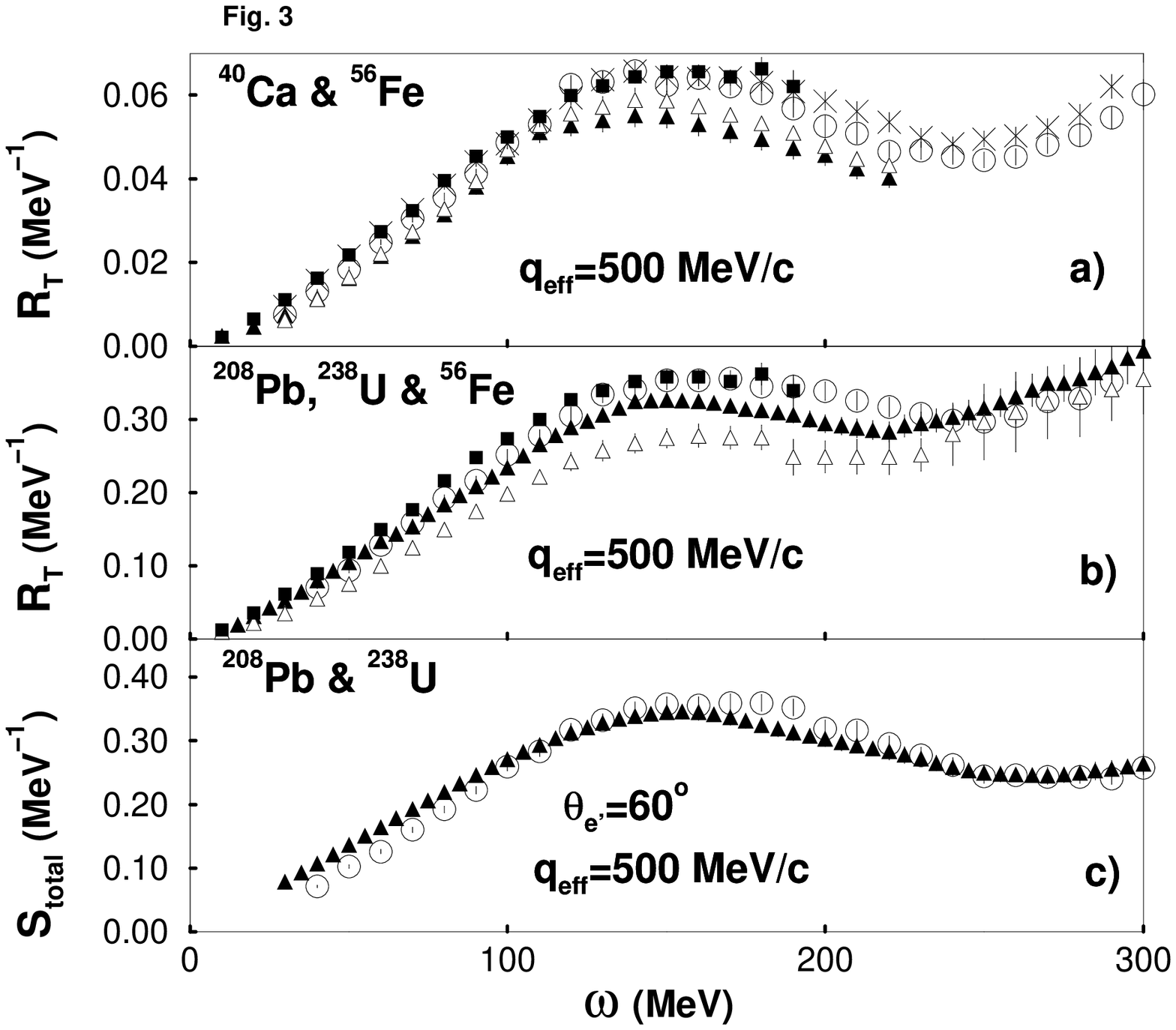,width=18.cm,height=18.cm}
\end{center}
\end{figure}

\newpage
\begin{figure}
\begin{center}
%\centerline{\includegraphics[scale=1]{fig-4-noir.eps}}
\epsfig{file=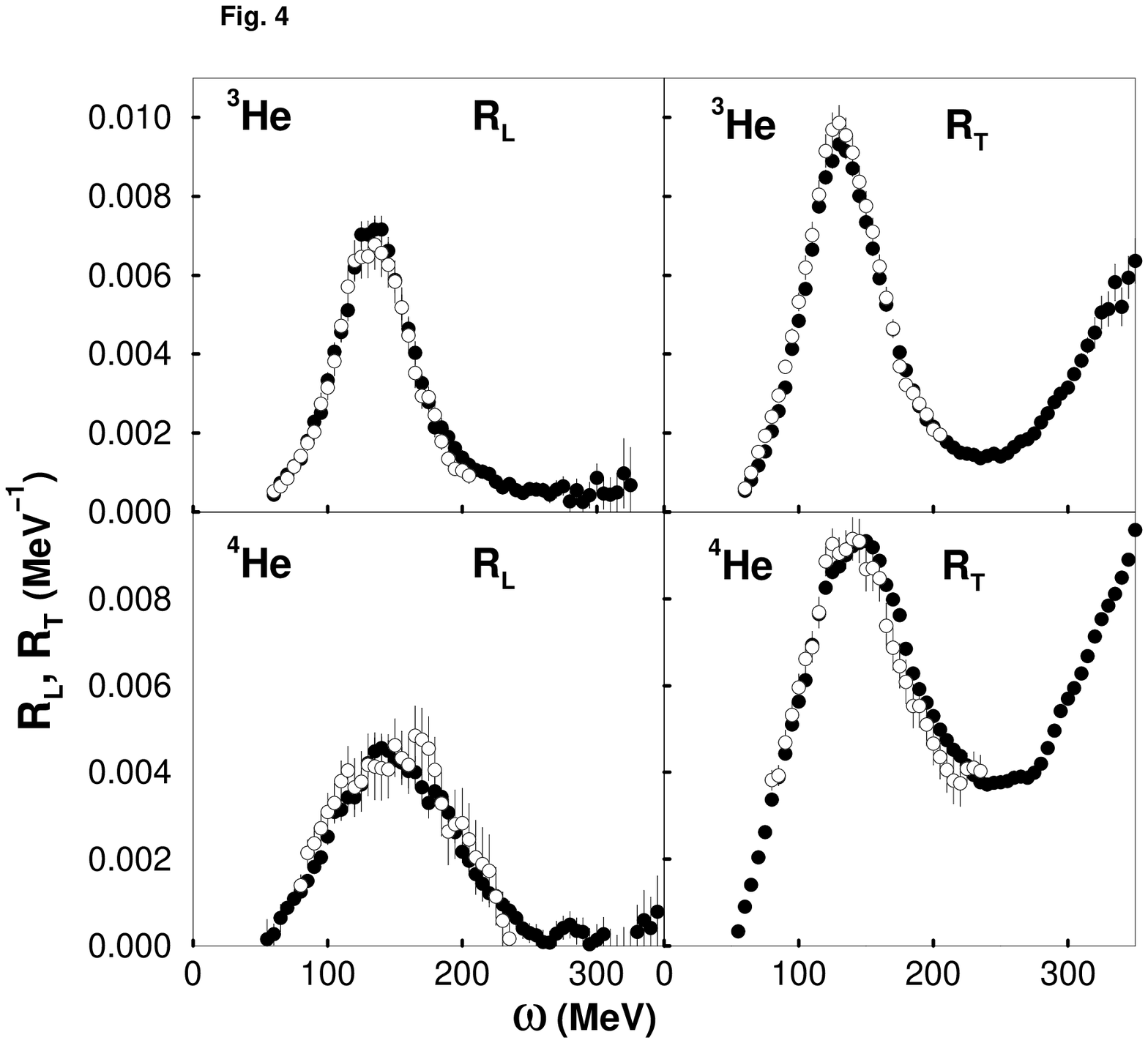,width=18.cm,height=18.cm}
\end{center}
\end{figure}

\newpage
\begin{figure}
\begin{center}
%\centerline{\includegraphics[scale=1]{fig-5-noir.eps}}
\epsfig{file=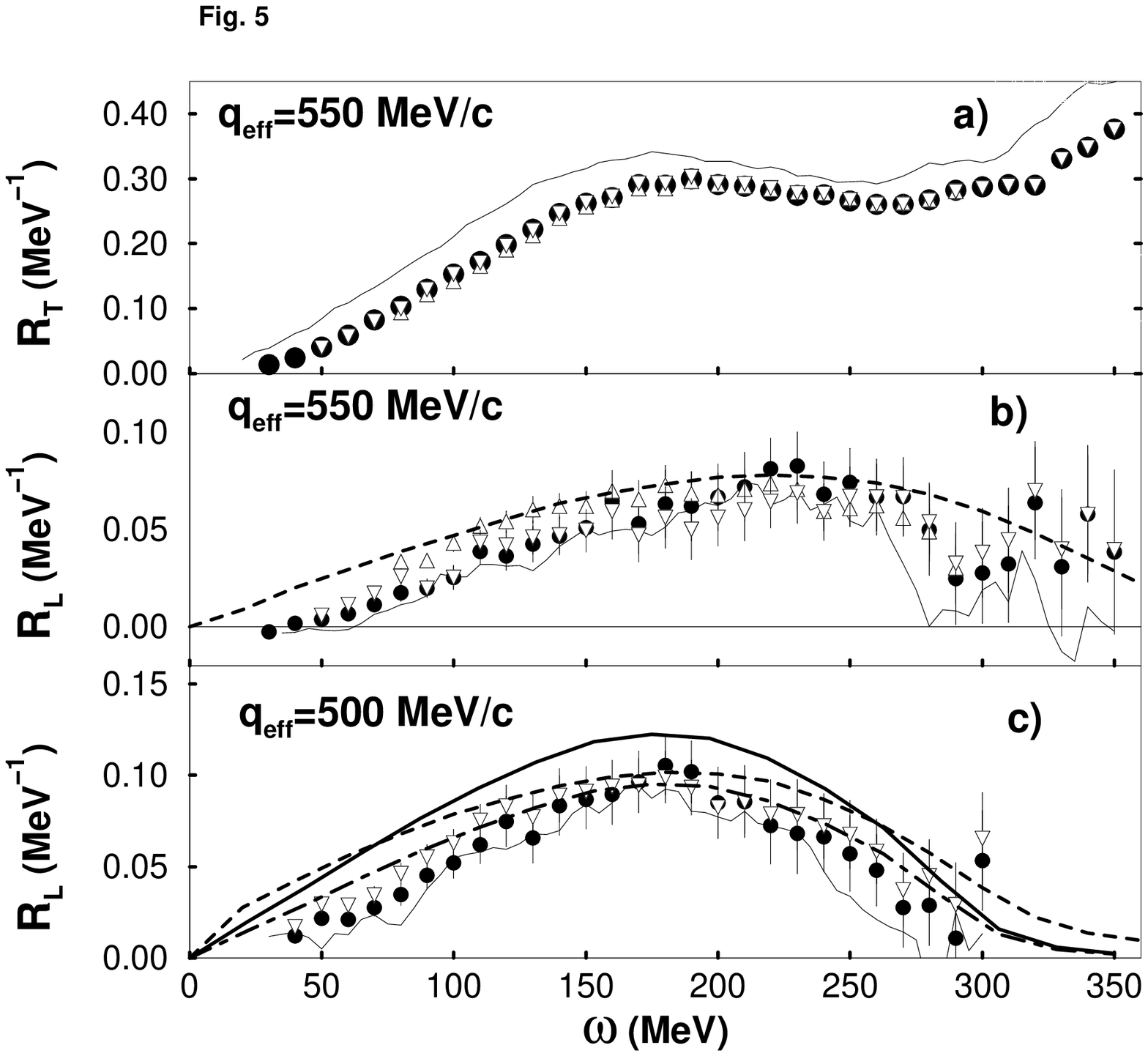,width=18.cm,height=18.cm}
\end{center}
\end{figure}

\newpage
\begin{figure}
\begin{center}
%\centerline{\includegraphics[scale=1]{fig-6-noir.eps}}
\epsfig{file=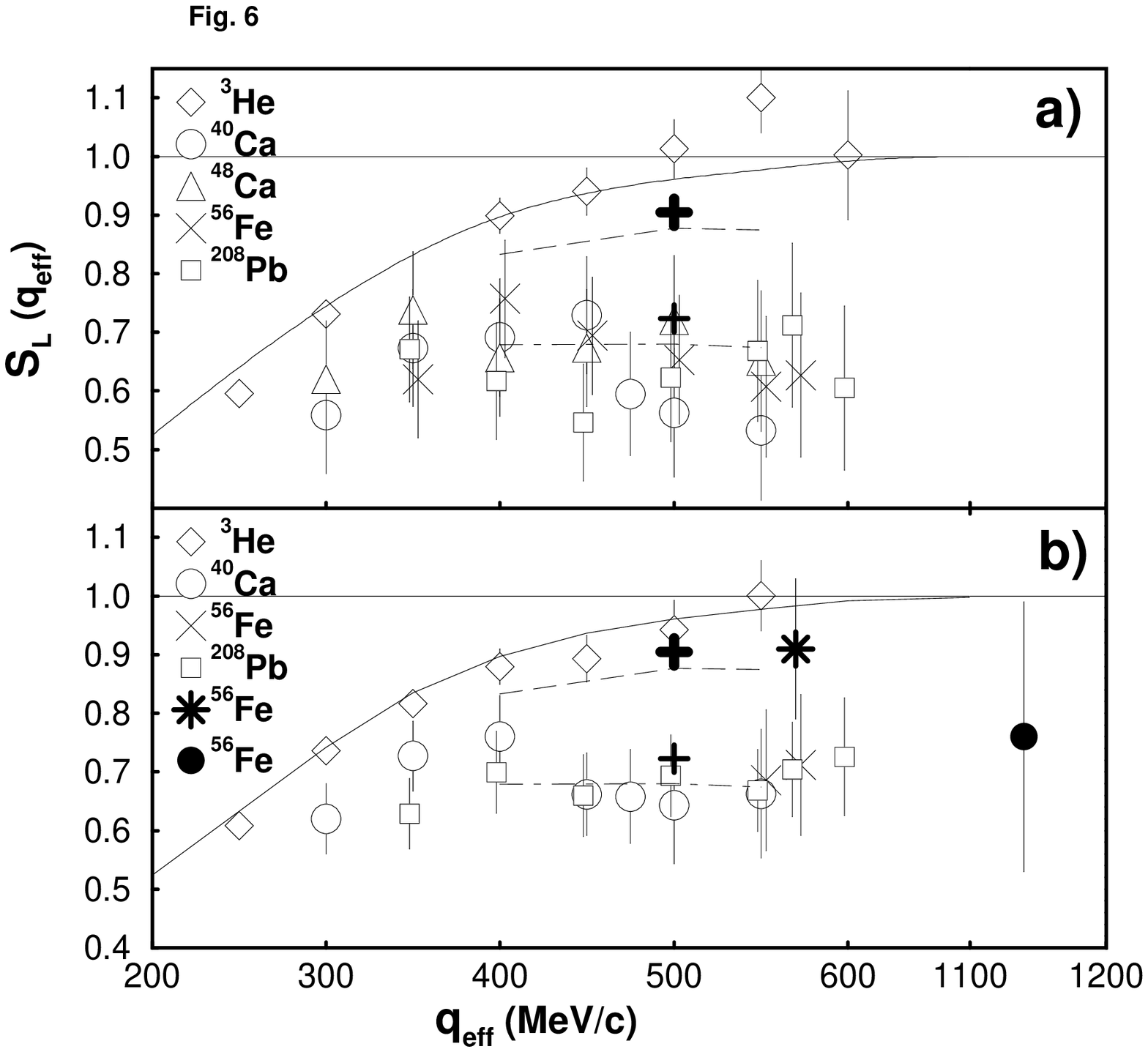,width=18.cm,height=18.cm}
\end{center}
\end{figure}

\end{document}